\documentclass[twocolumn,showpacs,preprintnumbers,amsmath,amssymb]{revtex4}
\usepackage{txfonts}
\usepackage{mathrsfs}
\usepackage{amssymb}

\usepackage{graphicx}
\usepackage{bm}

\begin{document}
\title{Security of "Counterfactual Quantum Cryptography"}
\author{Zhen-Qiang Yin, Hong-Wei Li, Zheng-Fu Han*, Guang-Can Guo}
\affiliation{Key Laboratory of Quantum Information\\ University of
Science and Technology of China\\ Hefei 230026\\ China}

\begin{abstract}
 Recently, a "counterfactual" quantum key distribution scheme was
proposed by Tae-Gon Noh \cite{N09}. In this scheme, two legitimate
distant peers may share secret keys even when the information
carriers are not traveled in the quantum channel. We find that this
protocol is equivalent to an entanglement distillation protocol
(EDP). According to this equivalence, a strict security proof and
the asymptotic key bit rate are both obtained when perfect single
photon source is applied and Trojan-horse attack can be detected. We
also find that the security of this scheme is deeply related with
not only the bit error rate but also the yields of photons. And our
security proof may shed light on security of other two-way
protocols.
\end{abstract}

\pacs{03.67.Dd}

\maketitle

 Quantum Key Distribution (QKD) \cite{BB84, Ekert1991,Gisin's review} can
enable two distant peers (Alice and Bob) to share secret random
string of bits, called key. With QKD and one-time-pad, unconditional
secure communication is possible. The most commonly used QKD
protocol is BB84, in which Alice encodes the state of a single
photon, transmits it to Bob through a quantum channel which is
accessed by a eavesdropper Eve, finally Bob projects this photon
into some states. Not only the BB84 protocol, nearly all of QKD
protocols must transmit information carriers (usually, a single
photon) in a public quantum channel. Many successful experiments of
QKD \cite{qkd1, qkd2, qkd3, qkd4, qkd5, qkd6, qkd7} have been
achieved during the past decade.

 Quite interestingly, Tae-Gon Noh proposed a QKD protocol (N09) \cite{N09}, in which
the distribution of a secret key bit can be accomplished even though
a photon carrying secret information is not in fact transmitted
through the quantum channel. Let us introduce the process of N09
protocol briefly.

 In N09 protocol, Alice randomly encodes single photon horizontal-polarized state
$|H\rangle$ as bit 0 or vertical-polarized state $|V\rangle$ as bit
1 and then inputs this photon to the port $2$ of a beam-splitter
(BS), whose the reflection and transmission modes are written as $a$
and $b$ respectively. For example, if Alice emits $|H\rangle$, the
quantum state of this photon will be
$|\psi_{H(V)}\rangle=(i|H(V)\rangle_a|0\rangle_b+|0\rangle_a|H(V)\rangle_b)/\sqrt{2}$,
in which we consider a $\pi/2$ phase is always added to reflection
case and there's no phase change to transmission mode. The key point
is that mode $a$ is kept by Alice, while mode $b$ represents the
quantum channel between Alice and Bob. Thus, Eve can only access the
mode $b$, while mode $a$ is unaffected by Eve. Bob will choose to
detect the $|H\rangle_b$ by his single photon detector (SPD) $D_3$
and just reflect other components of mode $b$ as bit 0 or detect the
$|V\rangle_b$ through $D_3$ and just reflect other components of
mode $b$ as bit 1. This operation can be viewed as a random
projection to $|X\rangle_b\langle X|$, which will be detected by the
detector $D_3$ and $1-|X\rangle_b\langle X|$, in which $X=H$ or
$X=V$. Bob's operation can be implemented by optical switches and
polarization-beam-splitter (PBS). To detect the intrusion of Eve,
Alice and Bob may compare the initial polarization state and the
detected polarization state, if $D_3$ clicks.

 The mode $b$ reflected by Bob will return to the Alice's BS and at
the same time the mode $a$ will also arrive at this BS due to the
reflection by a mirror owned by Alice. If the bit choices of Alice
and Bob are different, then the photon will output from the port $2$
of Alice's BS and then hit Alice's SPD $D_2$ due to the quantum
interference. Conversely, if the bit choices are the same, Bob will
get a click in $D_3$ with probability $1/2$, which means the photon
was at mode $b$. But, with another probability of $1/2$, the photon
is at mode $a$ and thus Bob get no click in $D_3$, Alice will get
one click in $D_2$ or $D_1$ with equal chances. Therefore, a click
from $D_1$ means the generation of one bit secret key. The clicks of
$D_1$ can only step from the photon at mode $a$ not the quantum
channel mode $b$. Thus we say in N09 the task of distributing secret
key bit can be finished when the information carriers are not
traveled in quantum channel.

 The security of N09 has not been proved though there are some discussion
on particular attacks. The security of this protocol cannot be
followed by the claim that Eve cannot access the whole information
carriers. Although some simple attacks such as Eve detects the
polarization of mode $b$, will spoil the quantum interference and
introduce bit error rate of key bits. Eve may entangle her ancilla
with the information carrier and apply different operations
to the go and return mode $b$. Eve is able to get some bit keys
without introducing bit error. It's totally different with BB84
protocol, which Eve cannot launch an effective attack without
introducing bit error in ideal case. Thus a strict security proof is
in urgent need for N09 protocol.

 In this paper, we put forward a security proof of N09 protocol when Trojan-horse-like attack \cite{trojan} is prohibited.
We find that the security of N09 is highly related to not only the
bit error rate of key, but also the counting rates of $D_1$ and
$D_2$. Inspired by Ref\cite{shor and preskill}, we propose a
entanglement distillation protocol (EDP) which is totally equivalent
to the N09 protocol. Here, the meaning of this equivalence between
the two protocols is: to Alice and Bob, the generated secret key is
the same; to Eve, the available information is also the same. The
EDP is illustrated in Fig.1 and the detailed steps are as follows:

\begin{figure}[!h]
\center \resizebox{6.5cm}{!}{\includegraphics{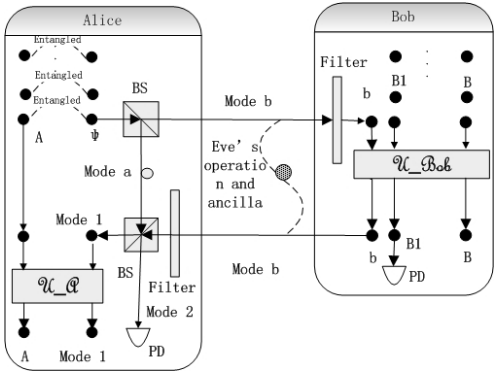}}
\caption{A and $\psi$ represent Alice's initial entangled particles; BS: beam-splitter; 
Filter: quantum operation controlled by Alice or Bob, which can project mode $b$ into the Hilbert space 
spanned by $|0\rangle$, $|H\rangle$, and $|V\rangle$. And a failure of this filtering operation results in the abortion 
of the whole protocol; B1 and B represent Bob's initial particles; PD: polarization detector which detects particles 
with projectors $|0\rangle\langle 0|$, $|H\rangle\langle H|$, and $|V\rangle\langle V|$.}
\end{figure}

 (1). Alice prepares $N$ pairs of entanglement states
$|\Psi\rangle_A=(|H\rangle_A|\psi_H\rangle+|V\rangle_A|\psi_V\rangle)/\sqrt{2}$,
in which, the particle $A$ and mode $a$ is protected in Alice's
security zone, while mode $b$ is the quantum channel between Alice
and Bob. Bob also prepares $N$ pairs of states
 $|\Psi\rangle_B=(|H\rangle_B+|V\rangle_B)|0\rangle_{B1}/\sqrt{2}$,
in which, the particles $B$ and $B1$ are all ancilla owned
by Bob, and Eve has no chance to access them. Alice sends all of the
modes $b$ of the $N$ pairs of entanglement states and announces this fact publicly.

 (2). After passing through the quantum channel controlled by Eve, the mode $b$ of $n$th $|\Psi\rangle_A$ will
enter Bob's security zone. Bob will first project the mode $b$ with
projectors $|0\rangle\langle 0|+|H\rangle\langle H|+|V\rangle\langle
V|$, and $I-|0\rangle\langle 0|-|H\rangle\langle H|-|V\rangle\langle
V|$. If Bob detects the mode $b$ through
the projective measurement $I - |0><0| - |H><H| - |V><V|$, he will
abort the protocol. This operation is carried out by filter in Bob' security zone as in Fig.1. If not, Bob will apply an unitary transformation
$\mathscr{U}_{Bob}$ to this mode b and particle $B$ and $B1$ of
$n$th $|\Psi\rangle_B$. $\mathscr{U}_{Bob}$ is defined as:
$\mathscr{U}_{Bob}|H\rangle_{B}|0\rangle_{B1}|0\rangle_b=|H\rangle_{B}|0\rangle_{B1}|0\rangle_b$,
$\mathscr{U}_{Bob}|H\rangle_{B}|0\rangle_{B1}|H\rangle_b=|H\rangle_{B}|H\rangle_{B1}|0\rangle_b$,
$\mathscr{U}_{Bob}|H\rangle_{B}|0\rangle_{B1}|V\rangle_b=|H\rangle_{B}|0\rangle_{B1}|V\rangle_b$,
$\mathscr{U}_{Bob}|V\rangle_{B1}|0\rangle_{B1}|0\rangle_b=|V\rangle_{B}|0\rangle_{B1}|0\rangle_b$,
$\mathscr{U}_{Bob}|V\rangle_{B1}|0\rangle_{B1}|H\rangle_b=|V\rangle_{B}|0\rangle_{B1}|H\rangle_b$,
$\mathscr{U}_{Bob}|V\rangle_{B1}|0\rangle_{B1}|V\rangle_b=|V\rangle_{B}|V\rangle_{B1}|0\rangle_b$.
After this transformation, Bob will detect the particle $B1$ with
projectors $|0\rangle\langle 0|$, $|H\rangle\langle H|$ and
$|V\rangle\langle V|$ and record the result. After that, the mode $b$
will re-enter the quantum channel.

 (3). After traveling along quantum channel controlled by Eve, the $n$th mode
$b$ will re-enter Alice's security zone. Before Alice combines this
mode $a$ and mode $b$ of $n$th $|\Psi\rangle_A$ in a BS at the same
time, Alice must apply the same projection as to Bob's projection in
step (2) to detect any possible Trojan-horse attack. This is done by filter in Alice's security zone as in Fig.1. Consider the
normal attenuation of mode $a$ is $\eta$, the effective state of
mode $a$ after this BS is
$|H(V)\rangle_a\longrightarrow\sqrt{\eta}(|H(V)\rangle_{1}+i|H(V)\rangle_{2})$.
For mode $b$,
$|H(V)\rangle_b\longrightarrow(|iH(V)\rangle_{1}+|H(V)\rangle_{2})/\sqrt{2}$.

 (4). For each trial, Alice measures the
mode $2$ with the following projectors: $|0\rangle_{22}\langle 0|$,
$|H\rangle_{22}\langle H|$, and $|V\rangle_{22}\langle V|$. This operation corresponds to the PD in Fig.1. If a
polarization state $H$ or $V$ of mode $2$ is observed by Alice, she
will measure the polarization of corresponding particle $A$ and the
result is recorded by her.
 If Alice gets $|0\rangle_2$ in her
measurement, Alice will detect if the polarization of mode $1$ and
the corresponding particle $A$ is the same. This operation can be
done by unitary transformation defined by
$\mathscr{U}_A|H(V)\rangle_A|0\rangle_1|a_0\rangle=|H(V)\rangle_A|0\rangle_1|a_0\rangle$,
$\mathscr{U}_A|H(V)\rangle_A|H\rangle_1|a_0\rangle=|H(V)\rangle_A|H\rangle_1|a_1(a_2)\rangle$,
$\mathscr{U}_A|H(V)\rangle_A|V\rangle_1|a_0\rangle=|H(V)\rangle_A|V\rangle_1|a_2(a_1)\rangle$,
and $|a_0\rangle$, $|a_1\rangle$ and $|a_2\rangle$ are all quantum states of Alice's
ancilla and orthogonal with each other. Now Alice detects the
$a$ with projectors $|a_0\rangle\langle a_0|$, $|a_1\rangle\langle
a_1|$ and $|a_2\rangle\langle a_2|$. If the output of Alice's
measurement on $a$ is $|a_1\rangle$, Alice will preserve the
corresponding particle $A$, $1$ for the following
process. And these $A$ and $1$ are called polarization consistent particles (PCPs).If Alice obtains $|a_2\rangle$, she measures the
polarization state of corresponding particles $1$ and $A$, which are called non-polarization-consistent particles (NPCPs) for abbreviation, and
records the results.

 (5). After the transmission of $N$ particles has completed, Bob tells Alice the results of detection of each
$B1$. Alice and Bob disregard all the particles corresponding to non-vacuum $B1$. Now, the following steps are only
carried out for the cases that $B1$ is in vacuum. 
Alice asks Bob to measure the polarization of
particles $B$ corresponding to NPCPs $A$.  And then Alice and Bob randomly select
half of the PCPs $A$, $1$ and its corresponding $B$, and measure them with the projectors
$|H\rangle\langle H|$ and $|V\rangle\langle V|$. Hence, the
probabilities $Pr(X_AY_B0_{B1}Z_{D})$, in which $X, Y, Z = H, V$ and
$D=1, 2$, are obtained by Alice and Bob.

 (6). According to all of the probabilities observed in step (5), Alice and Bob may carry
out EDP for the other half of the PCPs $A$, $1$ and its corresponding $B$.

 Since Eve cannot access Alice and Bob's ancillas, this
virtual entanglement protocol is equivalent to N09 from Eve's view.
To Alice and Bob, the key generated by the two protocols is totally
the same. Therefore, the security analysis of N09 protocol can be
carried out on this EDP. On the other hand, the EDP can be reduced
to N09 with Shor and Preskill's method \cite{shor and preskill,
GLLP}.

 The initial state of Alice is given by:

\begin{equation}
\begin{aligned}
|\Psi_{ini}\rangle^{\otimes N}_{A}=
(\frac{1}{\sqrt{2}}|0\rangle_{Aa}|0\rangle_b+\frac{1}{2}|H\rangle_{Aa}|H\rangle_b+\frac{1}{2}|V\rangle_{Aa}|V\rangle_b)^{\otimes
N}
\end{aligned}
\end{equation}
,in which,
$|0\rangle_{Aa}=(i|H\rangle_A|H\rangle_a+i|V\rangle_A|V\rangle_a)/\sqrt{2}$,
$|H\rangle_{Aa}=|H\rangle_A|0\rangle_a$, and
$|V\rangle_{Aa}=|V\rangle_A|0\rangle_a$. We also define
$|0\rangle=(1,0,0)^T$, $|H\rangle=(0,1,0)^T$, and
$|V\rangle=(0,0,1)^T$.

 We must point out only mode $b$ can be input into Alice and Bob, and the state of any modes $b$
after Eve's operation must be in a Hilbert space spanned by
$|0\rangle$, $|H\rangle$ and $|V\rangle$ since any state out of the
Hilbert space may be detected by Bob and Alice's projection
$1-|0\rangle\langle 0|-|H\rangle\langle H|-|V\rangle\langle V|$,
which results in the abortion of the whole protocol. Above
assumptions justify the negligence of Trojan attack, which makes the
security of nearly all of "go and return" QKD protocols to be
inexplicit. The most general attack is that: firstly, Eve may apply
an unitary transformation $\mathscr{U}_{Eve}$ to all the N $b$ modes
and her ancilla $e$. Particularly, we consider the evolution
of $l$th communication. This step can be described mathematically
like this:

\begin{equation}
\begin{aligned}
\mathscr{U}_{Eve}|\Psi_{ini}\rangle^{\otimes N}_{A}|e\rangle&=
\sum_{T(n\neq l)}(C_{T, T(l)=0}|T, T(l)=0\rangle_{Aa}\mathscr{U}_{Eve}|T, T(l)=0\rangle_b|e_0\rangle\\
&+C_{T, T(l)=H}|T, T(l)=H\rangle_{Aa}\mathscr{U}_{Eve}|T, T(l)=H\rangle_b|e_0\rangle\\
&+C_{T, T(l)=V}|T, T(l)=V\rangle_{Aa}\mathscr{U}_{Eve}|T,
T(l)=V\rangle_b|e_0\rangle)
\end{aligned}
\end{equation}
in which, $T$ is a list like $t_1...t_n...t_N$, $t_n=0, H, V$, and
$C$ is constant. Consider any state
$|T=t_1....t_l...t_N\rangle_b|e_0\rangle$ must be transformed to a
superposition which consists of three classes: $t_l=0$, $t_l=H$ or
$t_l=V$. In the next step Bob applies $\mathscr{U}_{Bob}$ to the N
$b$ modes, $B$ and $B1$. The result of Bob's operation can be
re-written like this:

\begin{equation}
\begin{aligned}
&\mathscr{U}_{Bob}[\frac{1}{\sqrt{2}}(H_B+V_B)]^{\otimes
N}\mathscr{U}_{Eve}|\Psi_{ini}\rangle^{\otimes N}_{A}|e_0\rangle\\
&=\frac{1}{2}0^{(l)}_{Aa}\{\Gamma_{00}(H^{(l)}_B+V^{(l)}_B)0^{(l)}_{B1}0^{(l)}_b+\Gamma_{0H}(H^{(l)}_BH^{(l)}_{B1}0^{(l)}_b+V^{(l)}_B0^{(l)}_{B1}H^{(l)}_b)\\
&+\Gamma_{0V}(H^{(l)}_B0^{(l)}_{B1}V^{(l)}_b+V^{(l)}_BV^{(l)}_{B1}0^{(l)}_b)\}\\
&+\frac{1}{2\sqrt{2}}H^{(l)}_{Aa}\{\Gamma_{H0}(H^{(l)}_B+V^{(l)}_B)0^{(l)}_{B1}0^{(l)}_b+\Gamma_{HH}(H^{(l)}_BH^{(l)}_{B1}0^{(l)}_b+V^{(l)}_B0^{(l)}_{B1}H^{(l)}_b)\\
&+\Gamma_{HV}(H^{(l)}_B0^{(l)}_{B1}V^{(l)}_b+V^{(l)}_BV^{(l)}_{B1}0^{(l)}_b)\}\\
&+\frac{1}{2\sqrt{2}}V^{(l)}_{Aa}\{\Gamma_{V0}(H^{(l)}_B+V^{(l)}_B)0^{(l)}_{B1}0^{(l)}_b+\Gamma_{VH}(H^{(l)}_BH^{(l)}_{B1}0^{(l)}_b+V^{(l)}_B0^{(l)}_{B1}H^{(l)}_b)\\
&+\Gamma_{VV}(H^{(l)}_B0^{(l)}_{B1}V^{(l)}_b+V^{(l)}_BV^{(l)}_{B1}0^{(l)}_b)\}\\
\end{aligned}
\end{equation}
, in which $\Gamma$ represents the arbitrary state of all particles
of $n\neq l$ and Eve's ancilla.

  Thirdly, another unitary transformation $\mathscr{U}'_{Eve}$ will
be applied to all the modes $b$ and $\Gamma$ by Eve. We note that
$\mathscr{U'}_{Eve}$ is arbitrary, for example,
$\mathscr{U'}_{Eve}\Gamma_{XY}Z^{(l)}_b=\Gamma_{XYZ0}0^{(l)}_b+\Gamma_{XYZH}H^{(l)}_b+\Gamma_{XYZV}V^{(l)}_b$,
in which $X,Y,Z=0,H,V$. For simplicity, we consider the Alice's
detectors and Bob's detector never clicks twice in one
communication. This condition can be justified in practical cases,
due to the lower dark counts of SPD. Hence, we obtain
$\Gamma_{0H}=\Gamma_{0V}=0$ and
$\mathscr{U'}_{Eve}\Gamma_{00}0^{(l)}_b=\Gamma_{0000}0^{(l)}_b$. We
also define $|K\rangle$, $K=0,1,2...$ is a set of well-defined basis
for all $\Gamma$ states, and $C_K(ABCD)=\langle
K|\Gamma_{ABCD}\rangle$, $A,B,C,D=0,H,V$. According to above
assumptions we may give the density matrix for the $l$th particles
$A$, $B$, $B1$, and modes $a$ and $b$ in the following equation:

\begin{equation}
\begin{aligned}
\rho^{(l)}_{AB}&=\frac{1}{4}\sum_KP\{0_{Aa}[(H_B+V_B)0_{B1}C_K(0000)0_b]\\
&+\frac{1}{\sqrt{2}}H_{Aa}\sum_{x}[(H_B+V_B)0_{B1}C_K(H00x)x_b\\
&+H_BH_{B1}C_K(HH0x)x_b+V_B0_{B1}C_K(HHHx)x_b\\
&+H_B0_{B1}C_K(HVVx)x_b+V_BV_{B1}C_K(HV0x)x_b]\\
&+\frac{1}{\sqrt{2}}V_{Aa}\sum_{x}[(H_B+V_B)0_{B1}C_K(V00x)x_b\\
&+H_BH_{B1}C_K(VH0x)x_b+V_B0_{B1}C_K(VHHx)x_b\\
&+H_B0_{B1}C_K(VVVx)x_b+V_BV_{B1}C_K(VV0x)x_b]\}\\
\end{aligned}
\end{equation}
Here, $P\{X\}=|X\rangle\langle X|$ and $x$ in the summation notation
must be $0,H,V$. Note that the unitary of Eve's operation and the
assumption
$\mathscr{U'}_{Eve}\Gamma_{00}0^{(l)}_b=\Gamma_{0000}0^{(l)}_b$ must
result in $\sum_K|C_K(0000)|^2=1$.

 Now, the effective operation done by Alice can be described like $H(V)_a\rightarrow
\sqrt{\eta}(H(V)_1+iH(V)_2)/\sqrt{2}$ and $H(V)_b\rightarrow
(iH(V)_1+H(V)_2)/\sqrt{2}$.

 For simplicity, we
define the $\alpha^{(l)}_K=C_K(0000)$,
$\beta^{(l)}_K=iC_K(H00H)+iC_K(HVVH)$,
$\gamma^{(l)}_K=iC_K(H00V)+iC_K(HVVV)$,
$\beta'^{(l)}_K=iC_K(V00V)+iC_K(VHHV)$,
$\gamma'^{(l)}_K=iC_K(V00H)+iC_K(VHHH)$,
$\xi^{(l)}_K=iC_K(H00H)+iC_K(HHHH)$,
$\zeta^{(l)}_K=iC_K(H00V)+iC_K(HHHV)$,
$\xi'^{(l)}_K=iC_K(V00V)+iC_K(VVVV)$, and
$\zeta'^{(l)}_K=iC_K(V00H)+iC_K(VVVH)$. If Bob gets $|0\rangle_{B1}$
and Alice gets $|a_1\rangle$ in step (4) of the EDP, the sub-system
of A, B will be projected into:

\begin{equation}
\begin{aligned}
&\rho'^{(l)}_{AB1}\\
&=\frac{1}{\Lambda^{(l)}}\sum_{K}P\{H_AH_B(\sqrt{\eta}\alpha^{(l)}_K+\beta^{(l)}_K)+V_AV_B(\sqrt{\eta}\alpha^{(l)}_K+\beta'^{(l)}_K)\\
&+H_AV_B(\sqrt{\eta}\alpha^{(l)}_K+\xi^{(l)}_K)+V_AH_B(\sqrt{\eta}\alpha^{(l)}_K+\xi'^{(l)}_K)\}\\
\end{aligned}
\end{equation}
, where $\Lambda^{(l)}$ is normalization constance. Now, we can
analyze the bit error rate and phase error rate of
$\rho'^{(l)}_{AB1}$. Define
$|\phi^+\rangle_{AB1}=(H_AH_BH_1+V_AV_BV_1)/\sqrt{2}$,
$|\phi^-\rangle_{AB1}=(H_AH_BH_1-V_AV_BV_1)/\sqrt{2}$,
$|\psi^+\rangle_{AB1}=(H_AV_BH_1+V_AH_BV_1)/\sqrt{2}$, and
$|\psi^-\rangle_{AB1}=(H_AV_BH_1-V_AH_BV_1)/\sqrt{2}$, we can deduce
bit error rate $e^{(l)}_{bit}=_{AB1}\langle
\psi^+|\rho'^{(l)}_{AB1}|\psi^+\rangle_{AB1}+_{AB1}\langle
\psi^-|\rho'^{(l)}_{AB1}|\psi^-\rangle_{AB1}$ and phase error rate
$e^{(l)}_{ph}=_{AB1}\langle
\phi^-|\rho'^{(l)}_{AB1}|\phi^-\rangle_{AB1}+_{AB1}\langle
\psi^-|\rho'^{(l)}_{AB1}|\psi^-\rangle_{AB1}$.

 With the expression of $\rho^{(l)}_{AB}$ we can deduce the following probabilities for the $l$th communication:

\begin{equation}
\begin{aligned}
&2Pr^{(l)}(H_AV_B0_{B1}H_1)=\frac{1}{16}\sum_{K}|\sqrt{\eta}\alpha^{(l)}_K+\xi^{(l)}_K|^2\\
&Pr^{(l)}(H_AV_B0_{B1}H_2)=\frac{1}{16}\sum_{K}|\sqrt{\eta}\alpha^{(l)}_K-\xi^{(l)}_K|^2\\
&2Pr^{(l)}(V_AH_B0_{B1}V_1)=\frac{1}{16}\sum_{K}|\sqrt{\eta}\alpha^{(l)}_K+\xi'^{(l)}_K|^2\\
&Pr^{(l)}(V_AH_B0_{B1}V_2)=\frac{1}{16}\sum_{K}|\sqrt{\eta}\alpha^{(l)}_K-\xi'^{(l)}_K|^2\\
&2Pr^{(l)}(H_AH_B0_{B1}H_1)=\frac{1}{16}\sum_{K}|\sqrt{\eta}\alpha^{(l)}_K+\beta^{(l)}_K|^2\\
&Pr^{(l)}(H_AH_B0_{B1}H_2)=\frac{1}{16}\sum_{K}|\sqrt{\eta}\alpha^{(l)}_K-\beta^{(l)}_K|^2\\
&2Pr^{(l)}(V_AV_B0_{B1}V_1)=\frac{1}{16}\sum_{K}|\sqrt{\eta}\alpha^{(l)}_K+\beta'^{(l)}_K|^2\\
&Pr^{(l)}(V_AV_B0_{B1}V_2)=\frac{1}{16}\sum_{K}|\sqrt{\eta}\alpha^{(l)}_K-\beta'^{(l)}_K|^2\\
\end{aligned}
\end{equation}

 Recall that
$\sum_K|\alpha_K|^2=1$,
$\sum_K|\sqrt{\eta}\alpha^{(l)}_K+\beta^{(l)}_K|^2/16=2Pr^{(l)}(H_AH_B0_{B1}H_1)$,
and
$\sum_K|\sqrt{\eta}\alpha^{(l)}_K-\beta^{(l)}_K|^2/16=Pr^{(l)}(H_AH_B0_{B1}H_2)$,
we obtain
$\beta^{(l)}=\sum_K|\beta^{(l)}_K|^2=8(2Pr^{(l)}(H_AH_B0_{B1}H_1)+Pr^{(l)}(H_AH_B0_{B1}H_2))-\eta$.
By the same way, we obtain
$\beta'^{(l)}=\sum_K|\beta'^{(l)}_K|^2=8(2Pr^{(l)}(V_AV_B0_{B1}V_1)+Pr^{(l)}(V_AV_B0_{B1}V_2))-\eta$.
Thanks to Cauchy's inequality,
$(\sqrt{\sum_K|a_K|^2}-\sqrt{\sum_K|b_K|^2})^2\leqslant
\sum_K|a_K+b_K|^2\leqslant
(\sqrt{\sum_K|a_K|^2}+\sqrt{\sum_K|b_K|^2})^2$ always holds for
arbitrary complex numbers $a_K$ and $b_K$. Due to
$\sum_K|\xi^{(l)}_K-\xi'^{(l)}_K|^2=\sum_K|\sqrt{\eta}\alpha^{(l)}_K+\xi^{(l)}_K-\sqrt{\eta}\alpha^{(l)}_K-\xi'^{(l)}_K|^2/4$,
we obtain the upper bound of $\sum_K|\xi^{(l)}_K-\xi'^{(l)}_K|^2$ is
$\xi^{(l)}=8(\sqrt{Pr^{(l)}(H_AV_B0_{B1}H_1)}+\sqrt{Pr^{(l)}(V_AH_B0_{B1}V_1)})^2$.
With these parameters, $e^{(l)}_{ph}$ can be given by:

\begin{equation}
\begin{aligned}
e^{(l)}_{ph}&=\frac{1}{2\Lambda^{(l)}}\sum_K(|\beta^{(l)}_K-\beta'^{(l)}_K|^2+|\xi^{(l)}_K-\xi'^{(l)}_k|^2)\\
&\leqslant
\frac{1}{2\Lambda^{(l)}}((\sqrt{\beta^{(l)}}+\sqrt{\beta'^{(l)}})^2+\xi^{(l)})
\end{aligned}
\end{equation}

 Though $e^{(l)}_{ph}$ has been given, we cannot give the overall
$e_{ph}$ since $e^{(l)}_{ph}$ may be arbitrary correlated with
previous $l-1$ events. Thanks to Azuma's inequality \cite{Azuma,
Boileau}, for sufficient large $N$ pairs of $A$, $B$ and $1$,
differs between $e_{ph}$ and $\sum^N_{l=1}e^{(l)}_{ph}/N$ are
arbitrary small. Therefore, we obtain the overall phase error rate
$e_{ph}=\sum^N_{l=1}e^{(l)}_{ph}/N$.

 Also according to Azuma's inequality, we have
$\beta\triangleq\sum^N_{l=1}\beta^{(l)}/N=8(2Pr(H_AH_B0_{B1}H_1)+Pr(H_AH_B0_{B1}H_2))-\eta$,
$\beta'\triangleq\sum^N_{l=1}\beta'^{(l)}/N=8(2Pr(V_AH_B0_{B1}V_1)+Pr(V_AH_B0_{B1}V_2))-\eta$,
and $\sum^N_{l=1}\xi^{(l)}/N\leqslant
8(\sqrt{Pr(H_AV_B0_{B1}H_1)}+\sqrt{Pr(V_AH_B0_{B1}V_1)})^2\triangleq\xi$
always hold when $N$ is sufficient large. Recall
$\sum_K|\alpha^{(l)}_K|^2=1$, we obtain
\begin{equation}
\begin{aligned}
\Lambda^{(l)}=&\sum_K(|\sqrt{\eta}\alpha^{(l)}_K+\beta^{(l)}_K|^2+|\sqrt{\eta}\alpha^{(l)}_K+\beta'^{(l)}_K|^2\\
&+|\sqrt{\eta}\alpha^{(l)}_K+\xi^{(l)}_K|^2+|\sqrt{\eta}\alpha^{(l)}_K+\xi'^{(l)}_K|^2)\\
&\geqslant
(\sqrt{\eta}-\sqrt{\beta^{(l)}})^2+(\sqrt{\eta}-\sqrt{\beta'^{(l)}})^2
\end{aligned}
\end{equation}
Therefore, the overall phase error rate can be bounded through the
following inequality:

\begin{equation}
\begin{aligned}
e_{ph}&=\sum^N_{l=1}e^{(l)}_{ph}/N\\
&\leqslant
\frac{1}{N}\sum^{N}_{l=1}min\{\frac{(\sqrt{\beta^{(l)}}+\sqrt{\beta'^{(l)}})^2+\xi^{(l)}}{2((\sqrt{\eta}-\sqrt{\beta^{(l)}})^2+(\sqrt{\eta}-\sqrt{\beta'^{(l)}})^2)},1\}\\
&\leqslant
\frac{1}{N}\sum^{N}_{l=1}[min\{\frac{\beta^{(l)}}{(\sqrt{\eta}-\sqrt{\beta^{(l)}})^2},1\}+min\{\frac{\beta'^{(l)}}{(\sqrt{\eta}-\sqrt{\beta'^{(l)}})^2},1\}\\
&+min\{\frac{\xi^{(l)}}{4(\sqrt{\eta}-\sqrt{\beta^{(l)}})^2},1\}+min\{\frac{\xi^{(l)}}{4(\sqrt{\eta}-\sqrt{\beta'^{(l)}})^2},1\}]
\end{aligned}
\end{equation}
, in which $min\{x,y\}$ equals to the smaller one of $x$ and $y$.
Now the final problem is how to calculate the upper bound of $e_{ph}$
with constraints $\beta=\sum^N_{l=1}\beta^{(l)}/N$,
$\beta'=\sum^N_{l=1}\beta'^{(l)}/N$ and
$\xi=\sum^N_{l=1}\xi^{(l)}/N$. Note that $min\{x/(\sqrt{\eta}-\sqrt{x})^2, 1\}$ is a nonconvex
function about $x$ ($x=\beta^{(l)}, \beta'^{(l)}$). And it's easy to verify that $\sum^{N}_{l=1}min\{\xi^{(l)}/4(\sqrt{\eta}-\sqrt{\beta^{(l)}})^2,1\}$
will be maximized when all the denominators are equal.
Hence, we can obtain the following upper
bound of $e_{ph}$:

\begin{equation}
\begin{aligned}
e_{ph}\leqslant\frac{4\beta+4\beta'}{\eta}+\frac{\xi}{4(\sqrt{\eta}-\sqrt{\beta})^2}+\frac{\xi}{4(\sqrt{\eta}-\sqrt{\beta'})^2}
\end{aligned}
\end{equation}

  In fact, if there isn't Eve's attack, and no channel noises, Alice and Bob must find
$2Pr(H_AH_B0_{B1}H_1)=\eta/16$ and $Pr(H_AH_B0_{B1}H_2)=\eta/16$,
thus $\beta=0$. With the same way we obtain $\beta'=0$, $\xi=0$.
Thus pure maximal entanglement states
$(H_AH_BH_1+V_AV_BV_1)/\sqrt{2}$ can be shared between Alice and
Bob. Due to the equivalence of the N09 and EDP, we conclude that N09
is unconditional secure in the noiseless case. We must point out
that the unconditional security is under the assumption that Eve
cannot control the transmission efficiency of Alice's mode $a$ and
quantum efficiency of Alice and Bob's SPDs. This is different with
BB84, which is secure even if the efficiency of detectors are
controlled by Eve. 

 We also consider a typical noise channel case, in
which the visibility is $V$ and polarization flip probability when
photon flying in quantum channel is $p$. Then we maybe obtain
$Pr(H_AH_B0_{B1}H_1)=\eta/32$, $Pr(H_AH_B0_{B1}H_2)=\eta/16$,
$Pr(H_AV_B0_{B1}H_1)=(1-V)(1-p)\eta/16$, and
$Pr(V_AH_B0_{B1}V_1)=(1-V)(1-p)\eta/16$, from which we can deduce
the $e_{bit}=2(1-V)(1-p)/(1+2(1-V)(1-p))$ and $e_{ph}=(1-V)(1-p)/2$.
For example, let $V=0.98$, $p=0$, we find $e_{bit}=3.85\%$ while
$e_{ph}=1\%$. It's interesting that $e_{ph}$ may be smaller than
$e_{bit}$.

  In this paper, we have proved the unconditional security of N09 protocol
by considering its equivalence to a EDP process. According to Ref. \cite{Renner},
our security proof is also composable. Through estimating
the upper bound of the ${e_{ph}}$, we obtain the key bit rate. We
find the security of N09 protocol relies not only the bit error rate
but also some counting rates of SPDs. We must point out that our
security analysis is in an ideal situation, in which we assume that
perfect single photon source is applied, Alice and Bob can detect
any type of Trojan-horse attacks, the mode $a$'s evolution is
perfect and the efficiencies of SPDs are all constant. We believe
that our security analysis has given a solid foundation for
real-life N09. The possible lower phase error rate than bit error
rate may be an advantage of N09 protocol.

 This work was supported by National Fundamental Research Program of China
(2006CB921900), National Natural Science Foundation of China
(60537020,60621064) and the Innovation Funds of Chinese Academy of
Sciences. To whom correspondence should be addressed, Email:
zfhan@ustc.edu.cn.

\end{document}